\documentclass[useAMS,usegraphicx,usenatbib]{mn2e}
\usepackage{multirow}
\usepackage{mathtools}
\usepackage{mathabx}
\usepackage[usenames]{color}
\def\jref@jnl#1{{\rm#1}}

\def\actaa{\jref@jnl{Acta Astron.}}      
\def\aj{\jref@jnl{AJ}}                   
\def\araa{\jref@jnl{ARA\&A}}             
\def\apj{\jref@jnl{ApJ}}                 
\def\apjl{\jref@jnl{ApJ}}                
\def\apjs{\jref@jnl{ApJS}}               
\def\ao{\jref@jnl{Appl.~Opt.}}           
\def\apss{\jref@jnl{Ap\&SS}}             
\def\aap{\jref@jnl{A\&A}}                
\def\aapr{\jref@jnl{A\&A~Rev.}}          
\def\aaps{\jref@jnl{A\&AS}}              
\def\azh{\jref@jnl{AZh}}                 
\def\baas{\jref@jnl{BAAS}}               
\def\jrasc{\jref@jnl{JRASC}}             
\def\jcap{\jref@jnl{JCAP}}               
\def\memras{\jref@jnl{MmRAS}}            
\def\mnras{\jref@jnl{MNRAS}}             
\def\na{\jref@jnl{New Astronomy}}        
\def\pra{\jref@jnl{Phys.~Rev.~A}}        
\def\prb{\jref@jnl{Phys.~Rev.~B}}        
\def\prc{\jref@jnl{Phys.~Rev.~C}}        
\def\prd{\jref@jnl{Phys.~Rev.~D}}        
\def\pre{\jref@jnl{Phys.~Rev.~E}}        
\def\prl{\jref@jnl{Phys.~Rev.~Lett.}}    
\def\pasa{\jref@jnl{PASA}}               
\def\pasp{\jref@jnl{PASP}}               
\def\pasj{\jref@jnl{PASJ}}               
\def\qjras{\jref@jnl{QJRAS}}             
\def\rmxaa{\jref@jnl{Rev.Mex.AA}}       
\def\skytel{\jref@jnl{S\&T}}             
\def\solphys{\jref@jnl{Sol.~Phys.}}      
\def\sovast{\jref@jnl{Soviet~Ast.}}      
\def\ssr{\jref@jnl{Space~Sci.~Rev.}}     
\def\zap{\jref@jnl{ZAp}}                 
\def\nar{\jref@jnl{NewAR}}               
\def\nat{\jref@jnl{Nature}}              
\def\iaucirc{\jref@jnl{IAU~Circ.}}       
\def\aplett{\jref@jnl{Astrophys.~Lett.}} 
\def\apspr{\jref@jnl{Astrophys.~Space~Phys.~Res.}}
\def\bain{\jref@jnl{Bull.~Astron.~Inst.~Netherlands}} 
\def\fcp{\jref@jnl{Fund.~Cosmic~Phys.}}  
\def\gca{\jref@jnl{Geochim.~Cosmochim.~Acta}}   
\def\grl{\jref@jnl{Geophys.~Res.~Lett.}} 
\def\jcap{\jref@jnl{JCAP}}      %
\def\jcp{\jref@jnl{J.~Chem.~Phys.}}      
\def\jgr{\jref@jnl{J.~Geophys.~Res.}}    
\def\jqsrt{\jref@jnl{J.~Quant.~Spec.~Radiat.~Transf.}}
\def\memsai{\jref@jnl{Mem.~Soc.~Astron.~Italiana}}
\def\nphysa{\jref@jnl{Nucl.~Phys.~A}}   
\def\physrep{\jref@jnl{Phys.~Rep.}}   
\def\physscr{\jref@jnl{Phys.~Scr}}   
\def\planss{\jref@jnl{Planet.~Space~Sci.}}   
\def\procspie{\jref@jnl{Proc.~SPIE}}   

\title[High-energy neutrino fluxes from AGN]
{High-energy neutrino fluxes from AGN populations inferred from X-ray surveys}

\author[I. B. Jacobsen et al.]
{Idunn B.\,Jacobsen$^1$\thanks{%
E-mail: idunn.jacobsen.09@ucl.ac.uk (IBJ);
\newline \hspace*{1.2cm}  kinwah.wu@ucl.ac.uk (KW)},
Kinwah Wu$^1$\footnotemark[1],
Alvina Y.\,L.\,On$^1$%
,
Curtis J.\,Saxton$^{2}$ \vspace*{0.2cm} \\
  $^1$Mullard Space Sicence Laboratory, University College London, Holmbury St Mary, Dorking RH5 6NT\\
  $^2$Physics Department, Technion $-$ Israel Institute of Technology, Haifa 32000, Israel} 

\begin{document}

\date{date}

\pagerange{\pageref{firstpage}--\pageref{lastpage}} \pubyear{0000}

\label{firstpage}

\maketitle

\begin{abstract}
High-energy neutrinos and photons are complementary messengers,
   probing violent astrophysical processes and structural evolution of the Universe.
X-ray and neutrino observations jointly constrain conditions in active galactic nuclei (AGN) jets:
   their baryonic and leptonic contents, and particle production efficiency.
Testing two standard neutrino production models for local source Cen~A
   \citep{KT2008,BB2009},
   we calculate the high-energy neutrino spectra of single AGN sources
   and derive the flux of high-energy neutrinos expected for the current epoch.
Assuming that accretion determines both X-rays and particle creation,
   our parametric scaling relations predict neutrino yield in various AGN classes.
We derive redshift-dependent number densities of each class,
   from {\it Chandra} and {\it Swift}/BAT X-ray luminosity functions
   \citep{SGB2008,ACS2009}.
We integrate the neutrino spectrum expected from the cumulative history of AGN
   (correcting for cosmological and source effects, e.g. jet orientation and beaming).
Both emission scenarios yield neutrino fluxes well above limits set by {\it IceCube}
   (by $\sim 4$--$10^6 \times$ at 1~PeV, 
   depending on the assumed jet models for neutrino production).
This implies that:
  (i) Cen~A might not be a typical neutrino source as commonly assumed;
  (ii) both neutrino production models overestimate the efficiency;
  (iii) neutrino luminosity scales with accretion power differently among AGN classes and hence does not follow X-ray luminosity universally;
  (iv) some AGN are neutrino-quiet (e.g. below a power threshold for neutrino production);
  (v) neutrino and X-ray emission have different duty cycles (e.g. jets alternate between baryonic and leptonic flows);  or
  (vi) some combination of the above.
\end{abstract}


\section{Introduction} 

Astronomy has relied heavily on photon-based observations. Photons participate in electromagnetic interactions, and they inevitably suffer absorption and scattering within the emitting sources and in the media along the line of sight. Neutrinos are neutral, relativistic particles,  but, unlike photons, only interact weakly with matter. As they are practically unabsorbed and unscattered over a large distance, even propagating through very dense media, they can be used to probe the physics of systems at distances as far as the edge of the observable universe. Neutrinos are therefore complementary to photons as astrophysical messenger particles. 

Neutrinos can be generated in violent astrophysical environments. Active galactic nuclei (AGN) and the associated jets, together with stellar objects such as pulsars, magnetars, supernovae and $\gamma$-ray bursters, are identified as potential sources of high-energy neutrinos  \citep{BBM2005, W2007, B2008}. AGN are the most luminous persistent X-ray sources known. At their cores resides a massive black hole (with mass $M_{\bullet} \sim 10^6 - 10^9~{\rm M}_\Sun$), and the accretion of material into their central massive black hole powers the AGN activities. The accretion process in AGN is often accompanied by a material outflow, which manifests as relativistic jets at kpc to Mpc scales. 
Various scenarios for high-energy neutrino production in AGN jets have been proposed \citep[see e.g.][]{M1995,MREPS1999,AD2003,KT2008,BB2009}.  
The basic mechanism can be understood as follows. 
Charged hadrons, such as protons, are first accelerated to very high energies inside the jet. 
A possible acceleration site is at shocks formed inside the jet body \citep[e.g.][]{BR1974,H1979,BBR1984,BS1987}. 
The high-energy protons accelerated by the jet interact with the ambient particles 
  (e.g.\ cosmic microwave background (CMB) photons or the baryons in the environments), 
  which generates cascades of lighter children particles and subsequent production of charged pions ($\pi^\pm$ particles). 
The decay of these charged pions produces the high-energy neutrinos \citep[see e.g.][]{B2008, ABG2010}.    
Another possible acceleration site is at the jet base, where accretion inflow and relativistic outflow interact. 
The charged hadrons, presumably protons, are accelerated in shocks near the accretion disc 
  \citep[][see also \citealt{NMB1993,SS1996}]{SDSS1991}.   
Through proton-photon ($p\gamma$) interactions with the UV and X-ray photon fields from the accretion disc, 
  neutrinos are produced through the decay of pions. 

In a theoretical perspective, neutrino production is naturally associated with cosmic rays (CR), as high-energy neutrinos are products in the decay chain of energetic particles produced by interactions between CRs and ambient material and photons. In the neutral pion decay following the proton-proton ($pp$) and $p\gamma$ interactions, $\gamma$-ray emission is also produced at comparable energies. The remarkably detailed observed CR spectrum extends over eight orders of magnitude in energy \citep[e.g.][]{dermer2009,KO2011}, following a power-law with two clear breaks, and a suppression of flux towards the highest energies compatible with the GZK effect \citep{G1966,ZK1966}. The transition from a Galactic origin to an extragalactic origin is commonly assumed to occur around $4\times 10^{9}~\rm{GeV}$, considering the energetics of known Galactic sources and the non-correlation between local sources and CR events \citep[see e.g.][]{B2008,KO2011}. The highest energy CR events are therefore tracers of the acceleration processes within the sources, however due to cosmic magnetic fields, the CR particles lose directionality. The neutrinos and $\gamma$-rays produced within the sources are however not affected, and whereas $\gamma$-rays attenuate upon interaction with intergalactic media, neutrinos reach us virtually unimpeded. The CR spectrum hints that their sources might emit energetic neutrinos too. Studying neutrinos and $\gamma$-rays from these sources will enable an investigation of the accelerating region within the source itself. We take this connection a step further and relate the CR emission and its derivatives (neutrinos and $\gamma$-rays) to the accretion processes driving the AGN jet. 

Since CR, neutrino and $\gamma$-ray emissions are intrinsically linked, the CR and neutrino observations are thus complementary. One may naturally consider that the neutrino power scales with the $\gamma$-ray power of the AGN sources.
While this could be possible for individual sources, 
  the reality is more complicated when deriving a scaling relation 
  applicable to the whole AGN population collectively 
  or to an AGN subclass population 
  from the $\gamma$-ray observations.   
For instance, 
  by assuming that 10\% of the $\gamma$-ray background at the MeV energies is due to non-thermal emission 
  from AGN, such as Seyfert galaxies,    
  \cite{S2005} obtained a flux $\Phi_\nu \sim 10^{-18}~{\rm GeV^{-1}~cm^{-2}s^{-2}sr^{-2}}$ at 100~TeV for the $\mu$-neutrinos, 
  comparable the current flux limit 
  of $2.06^{+0.4}_{-0.3} \times 10^{-18} (E_\nu/100~{\rm TeV})^{-2.06\pm0.12}~{\rm GeV^{-1}~cm^{-2}s^{-2}sr^{-2}}$ 
  at the same energy set by the {\it IceCube} experiment \citep{AAA2015}.  
However, the prescription of \cite{S2005} gives a neutrino flux density higher
  by $\sim$$1.5$--$5$
  than the current observed limit 
  at PeV energies \cite[see fig. 12 in][]{AAA2015}. 
Moreover, it is unclear whether or not 10\% of the diffused MeV $\gamma$-rays observed in the sky is non-thermal emission from the Seyfert AGN and their relation to the neutrino generation process. 
It is also uncertain whether AGN in Seyfert galaxies are neutrino sources. 
Since the IceCube detection of TeV to PeV neutrinos \citep{AAA2013,IC2013,AAA2015}, consistent with an extragalactic origin, several studies have attempted to pinpoint the source class of these neutrino events \citep[see e.g.][]{H2014}. Using the photohadronic interaction channel the neutrino flux expected in blazars has been found to agree with the IceCube events assuming X-ray and $\gamma$-ray emission is produced through the $\pi^{0}$-decay \citep{krauss2015}. \citet{DMI2014} investigates the neutrino output by via the photohadronic channel, where the CR protons interact with internal or external radiation fields. They find that low-luminosity blazars are poor producers of neutrinos, whereas $\gamma$-ray bright flat spectrum radio quasar (FSRQ) blazars are promising candidates. In our work we explore the neutrino production efficiency in AGN populations by focusing on the common engine of the AGN power. X-ray and CR emission are both driven by the central accretion processes, and the latter will result in the emission of high-energy neutrinos.

In spite of decades of intense observational and theoretical studies, in particular in the radio and X-ray wavebands, 
there are still many outstanding questions regarding the dynamical and chemical properties of AGN and their jets. We are unsure how much mechanical energy is stored in a jet for given observed radio and/or X-ray luminosities \citep[see][]{WRB1999,MH2007,CB2009,SG2013} and we know little about the chemical ingredients in AGN jets \citep[see][]{LCB1992,WHO1998,GC2001,BRS2013}. 
In the context of neutrino production, we need first to 
  know whether AGN jets are predominantly baryonic, leptonic, 
  both baryonic and leptonic, or electromagnetic \citep[i.e.\ Poynting flux-dominated, see][]{L1976,LB1996,NTL2008}. 
We also need to know if baryons and leptons co-exist in the jet flow and 
 if AGN jets have alternating duty cycles of baryonic and leptonic flows, 
 analogous to active and dormant phases in terms of X-ray and radio emission.   

AGN are an inhomogeneous class of objects with diverse observational properties. 
For instance, they may be radio-loud (RL) or radio-quiet (RQ), 
 and their jets may be weak and episodic, or span a large spatial scale and continuously. 
There is no guarantee that neutrinos are produced in the same manner across all classes of AGN 
  and that all kinds of jets in AGN are equally efficient in neutrino emission.  
 
In this work, we address the above issues in the neutrino output in AGN jets, with an objective to set constraints on various scenarios of neutrino production in AGN, using a multimessenger approach, which combines the information obtained from X-ray observations and neutrino experiments. We use the X-ray survey observations of AGN by {\it Chandra} \citep{SGB2008} and {\it Swift} \citep{ACS2009} and derive the populations of various AGN at different cosmological epochs. We next apply the neutrino production models and determine the neutrino emission from individual AGN. From this, we compute the energy spectra and the flux limits of the neutrinos generated by different AGN classes and accumulated throughout the history of the Universe. We then compare the flux limits to the detection limit set by the {\it IceCube} neutrino experiment \citep{AAA2014b}, which constrains the particle content and physical properties of AGN jets, and verifies the neutrino production models proposed for different AGN classes.     

This paper is organized as follows: 
Section \ref{sec:agn_neutrinos} discusses the argument for AGN as candidate neutrino sources and the mechanisms leading to neutrino emission in AGN environments. 
Section \ref{sec:nu_models} describes the two hadron-channel models for neutrino production in AGN.   
Section \ref{sec:agn_populations} outlines the two X-ray surveys that we use, along with a description of the X-ray luminosity function (XLF) prescriptions and the calculations of the various AGN populations we derive from this data. 
Section \ref{sec:nu_spectra} presents the high-energy neutrino spectra obtained by our calculations, and 
Section \ref{sec:implications_discussion_conclusion} concludes with the astrophysical implications of our results. 

\section{Neutrinos from AGN jets}\label{sec:agn_neutrinos}

\subsection{AGN as candidate neutrino sources}\label{ssec:agn_fam} 
\begin{figure}
\centering
\includegraphics[width=8.4cm]{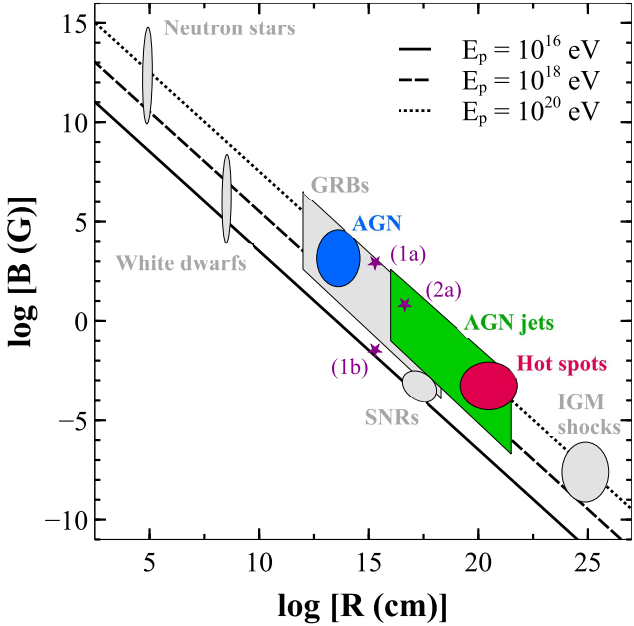}
\caption{Hillas diagram of the sources which are able to confine accelerated protons of maximum energies 
$E_{p,{\rm max}}=10^{16}$,  $10^{18}$  and $10^{20}~{\rm eV}$, with contours of various source candidates, adapted from fig.~8 in \citet{KO2011}. AGN and AGN jets meet the \citet{H1984} criterion for energetic protons, and therefore are strong candidates for the production of high-energy neutrinos. The three points denoted (1a), (1b) and (2a) refer to the location of Cen A on the Hillas plot with different considerations: (1a) follows the Hillas criterion (Equation~\ref{eqn:hillas_crit}) for proton confinement at $r=132 \ r_{\rm g}$, where the confinement is assumed according to \citet{RMR2011}; (1b) at the same location, but taking into account energy losses which lowers the maximal energy possible in the source, (2a) at a location of $r \approx 3000\ r_{\rm g}$, where confinement occurs according to \citet{BB2009} using the Hillas criterion. See Section \ref{sec:nu_models} for details.}
\label{fig:hillas_plot}
\end{figure}

Whether a charged particle could attain a certain energy depends on the duration of its confinement within the acceleration region. The \citet{H1984} criterion, which states  
\begin{equation}\label{eqn:hillas_crit}
E_{q, {\rm max}} = q B r  \ , 
\end{equation}
gives an estimate for the maximum energy $E_{q, {\rm max}}$ of a particle of charge $q$ accelerated in a region with a characteristic size $r$ and a magnetic field $B$.
AGN have emerged as candidate neutrino sources, due to their energetic nature and ability to accelerate charged particles to energies that facilitate the production of high-energy neutrinos. Taking that their outflows are hadronic, AGN and their jets are among a handful extragalactic source types that fit the requirements (see Fig.~\ref{fig:hillas_plot}), and hence are potential $10^{20}$~eV neutrino sources. 

AGN emit electromagnetic radiation (photons) over a broad waveband from radio, sub-mm, IR, optical, UV to X-ray and $\gamma$-ray. They are known as strong X-ray emitters, and many AGN are discovered by X-ray observations. In practice, compact extragalactic sources observed with a persistent X-ray luminosity above $L_{\rm X}\geq10^{42}~{\rm erg \ s}^{-1}$ could be safely assumed as an AGN.   

We would expect variations in the neutrino production rates from various AGN classes (Table \ref{table:agn_class}). Often AGN are categorized into various subtypes according to their observational properties at particular wavelengths. A common divide is luminosity in radio \citep[e.g.][]{A1993}, which depends on whether an AGN is bright in radio emission. It can be seen releasing its energy in two oppositely directed, highly collimated, relativistic jets, perpendicular to the accretion disc (RL AGN), or with no discernible jet structure (RQ AGN). The fraction of RL systems is about 10--20 per cent of the total AGN population \citep[e.g.][]{KSS1989,UP1995}, and the RL fraction is estimated to reach up to 50 per cent for quasars measured in X-rays \citep{dCLM1994}. 

RL systems are particularly important in the study of neutrino output in AGN. Their jets must consist of highly energetic, charged particles in order to produce their observed radio luminosities. Here, we present a brief review of the various RL AGN subclasses. A common classification scheme includes the orientation and brightness distribution of their jets \citep[e.g.][]{B1989, UP1995, T2008}. Radio galaxies (RGs) are observed when the jet has a viewing angle close to $90^\circ$. The active nucleus in these systems is fully obscured or partially obscured. RGs are generally separated into two Fanaroff-Riley (FR) types, distinguishable by the strength of their radio emission \citep[][]{FR1974}. FR Type I (FR~I) galaxies are of lower radio luminosity. They usually show a bright jet at the centre. FR Type II (FR~II) galaxies are more radio-luminous. They have relatively faint central jets, but with bright termination shocks at the tip of the jet-blown lobes. 

Radio-loud quasars (RLQs) and blazars are unobscured systems in which the jets are aligned along our line of sight or close to our line of sight. Their emission is therefore relativistically beamed. Radio quasars are among the very brightest and the most distant objects that we observe. They may be separated into lobe emission dominated systems with a steep radio spectrum (SSRQ) at higher viewing angles, or core emission dominated systems with a flat radio spectrum (FSRQ) at smaller viewing angles. Thus, the subclasses of radio quasars are also distinguished by the jet orientations. If the viewing angle is very small, the jet will be directed into our line of sight. This occurs in a blazar. The fraction of blazars is no more than 5\% of the total AGN population. Blazars can be categorized into the high-luminosity FSRQs and the low-luminosity BL Lacs. In the framework of the AGN unification model \citep{UP1995}, these two subclasses are intrinsically considered FR type II/RLQ and FR~I, respectively, where the jets are aligned in our line of sight direction. 

\begin{table}
\caption{\textbf{Classification of radio-loud AGN} adapted from \citet{UP1995}. The three classes of AGN are distinguished by inclination of the radio jet to our line of sight. RGs are at high viewing angles, and consist of low radio-luminous FR~Is and higher radio-luminous FR~IIs. At lower viewing angles, we find RLQs, with SSRQ to FSRQ for decreasing viewing angles. At the smallest viewing angles, where the jet is directly in our line of sight, we categorize blazars, with lower luminosity BL Lacs and higher luminosity FSRQs. The RLQs and blazars are all observed with beamed luminosities, and there exists a unification scheme across the three types (see the text).}
\label{table:agn_class}
\begin{tabular}{llll}
\hline
{Type:} & RG & RLQ & Blazars\\
\hline
${L_{\rm X} [{\rm erg \ s}^{-1}]}$: 	& $10^{42} - 10^{47}$ 	& $10^{44} - 10^{48.5}$ 	& $10^{44} - 10^{48.5}$ \\
\multirow{2}{*}{subclasses:} 		& FR~I 				& SSRQ 				& BL Lac \\
							& FR~II 				& FSRQ 				& FSRQ \\
\hline
\end{tabular}
\end{table}
%

\subsection{Neutrino production in AGN jets}\label{ssec:agn_nu}
In the hadronic scenario, energetic protons are source particles for neutrino production. Two interaction channels are proposed for neutrino production in AGN environments: one with protons interacting with ambient photons (photons from the accretion disc, synchrotron photons emitted in the jet, CMB photons strayed into the jet); and another with protons interacting with other protons within the jet or with protons of the external material entrained into the jet flow  \citep[see e.g.][and references therein]{E1979,MB1989,BRS1990,MREPS1999}. In the $p\gamma$ channel, pions are produced via 
\begin{eqnarray} \label{eqn:p-gamma-channel}
	p+\gamma  & \longrightarrow &  \Delta^+ \  \longrightarrow \  \left\{
		\begin{array}{l l}
			p + \pi^0  \\ 
			n + \pi^+ \\ 
			\drsh n + \gamma \longrightarrow p + \pi^- 
		\end{array} 
	\right. \ . 
\end{eqnarray} 
The decay branching ratios of the Delta resonance $\Delta^+$ are such that two-thirds will follow the $\pi^0$--channel, and the remaining third will produce charged pions $\pi^\pm$. The $pp$ interaction also leads to pion production, i.e.\, $p\,p \longrightarrow \{\pi^0, \pi^+, \pi^-\}$. Radiation fields are expected to be strong at the base of the jet. In this paper, we consider only models with $p\gamma$ interactions and leave those with $pp$ interactions to a future study.  

Neutral pions will decay to $\gamma$-rays ($\pi^0 \longrightarrow \gamma\gamma$), however the decay of charged pions produces electrons and neutrinos,
\begin{eqnarray} \label{eqn:pos-pion-decay}
	\pi^+  & \longrightarrow &\mu^+ + \nu_\mu \nonumber  \\ 
  	& &     \drsh e^+ + \nu_{\rm e} + \bar \nu_\mu  \ ; 
\end{eqnarray}
\begin{eqnarray} \label{neg-pion-decay}
	\pi^- & \longrightarrow &\mu^- + \bar{\nu}_\mu  \nonumber \\ 
  	&  &   \drsh e^- + \bar \nu_{\rm e} + \nu_\mu \ .
\end{eqnarray} 
High-energy CRs are also products of the interactions, as escaping neutrons could undergo $\beta$-decays ($n \longrightarrow p\, e^-{\bar \nu}_e$), leading to emission of neutrinos.

For neutrinos resulting from pion decay, the ratio of neutrino flavours at source is $(\nu_{\rm e}: \nu_{\mu}: \nu_{\tau})=(1:2:0)$. Due to neutrino oscillations as they propagate through space, we expect the detected ratio at Earth as $(\nu_{\rm e}: \nu_{\mu}: \nu_{\tau})=(1:1:1)$ \citep[e.g.][]{B2008}. We follow this commonly accepted picture in our paper.

However, deviations due to energy dependences on the decay rates and the strength of the source magnetic field can lead to energy loss of muons before decay (muon damping). In this case, the source ratio is lowered to $(\nu_{\rm e}: \nu_{\mu}: \nu_{\tau})=(0:1:0)$, as the electron neutrinos from the muon decays are of much lower energy than the muon neutrinos produced through the more energetic pion decays. This gives the detected flavour distribution as $(\nu_{\rm e}: \nu_{\mu}: \nu_{\tau})=(1:1.8:1.8)$ \citep[see][]{KW2005,P2008}.

\section{High-energy neutrino production models}\label{sec:nu_models}
In this study we consider the model proposed by \citet{KT2008} and the model proposed by \citet{BB2009}. In both models, $p\gamma$ interaction is the dominant source process, and follows the standard picture of flavour distribution at observation as outlined above. The primary protons are accelerated through shocks in the jet, with a power-law energy spectrum. The neutrino flux is scaled by CR events detected by the Pierre Auger Observatory (PAO), assuming that the events are of AGN origin. 

\subsection{The Koers \& Tinyakov (KT) model}\label{ssec:kt_mod}
\begin{figure}
\centering
\includegraphics[width=8.4cm]{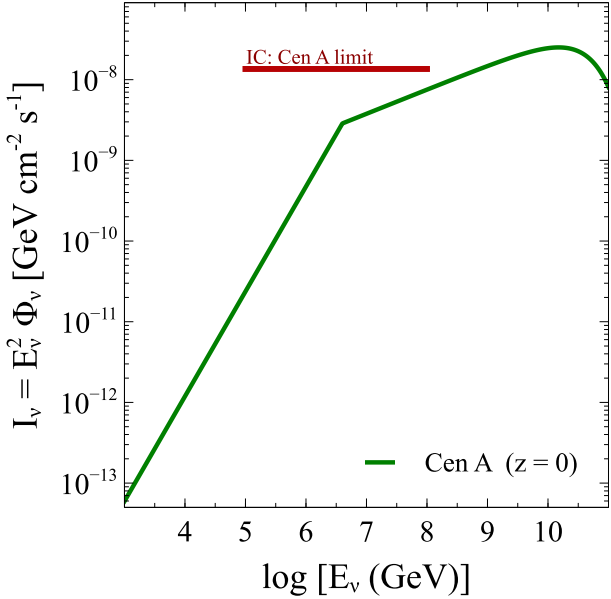}
\caption{Single source spectrum for Cen A, evaluated at a redshift $z=0$ \citep[CH;][]{KT2008}, compared with the upper flux limit for Cen A, determined by {\it IceCube} \citep{AAA2014a}.}
\label{fig:ktcena}
\end{figure}

The model by \citet{KT2008} studies the relation between diffuse and point-source neutrino emission, and uses the RG nearest to us, Cen A, as a typical source for neutrino production (Fig. \ref{fig:ktcena}). Cen A, which lies at a distance of about 3.4~Mpc \citep[e.g.][]{dV1979,SMW1996,I1998,EKW2004,HRH2010}, is an FR~I RG. As it is so close to us, it is also a well-used target for neutrino studies, as observations can be correlated with its location. The model thus follows a model of neutrino emission from Cen A by \citet[][CH]{CH2008}, and it is in turn based on work on a neutrino production model from extragalactic sources of hadronic origin by \citet{MPR2001}. Following the Hillas criterion, Cen A is a fairly good candidate for high-energy particle emission, however when accounting for energy losses, it falls within the boundary of maximum proton energy $E_{p,{\rm max}} \sim 10^{16}~{\rm eV}$ \citep{RMR2011}, and is from these considerations not able to produce the highest energy particles. The energy loss calculations assume an estimate for the mass of the central black hole $M_{\bullet} = 10^8\,{\rm M}_{\Sun}$, the location of the confinement and acceleration $R \sim 132~r_{\rm g}$ (where $r_{\rm g} = G M_{\bullet} c^{-2}$ is the gravitational radius of the central black hole), maximum proton energy $E_{p,{\rm max}}= 2 \times 10^7~{\rm GeV}$ and formulae for the evolution of the bulk Lorentz factor and magnetic field along the jet given in \citet[][see Fig. \ref{fig:hillas_plot}: point (1a) denotes the location of Cen A solely based on the Hillas criterion, whereas point (1b) shows where it lies if energy losses, with the dominant process being $p\, \gamma$ interactions, are taken into account]{RMR2011}.

\begin{figure*}
\begin{minipage}{15cm}
\centering
\resizebox{0.8\hsize}{!}{\includegraphics[angle=0]{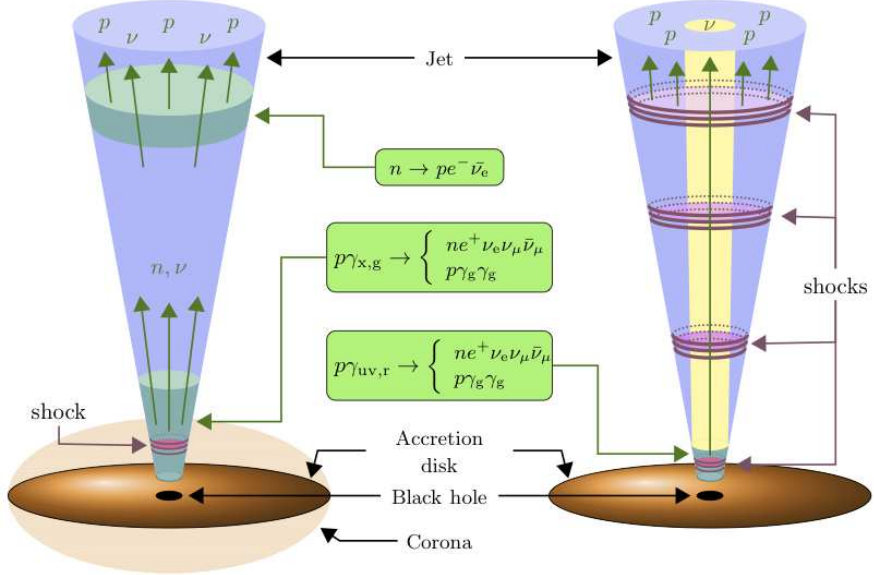}}  
\caption{Schematic illustration of the jet structure (not to scale). On the left is the KT model jet, where shocks at the base of the jet accelerate protons that subsequently interact with X-ray photons produced in inverse Compton processes in the corona. Neutrons and neutrinos escape the confining regions, however the neutrons suffer beta decays before leaving the jet, hence producing a population of CR protons, along with additional neutrinos. On the right is the BB model jet. At a few thousand gravitational radii, stable shocks accelerate protons that interact with the synchrotron photon field produced by relativistic electrons in the jet magnetic field. Neutrinos escape the jet in a collimated beam, whereas protons are continually accelerated along the jet, until they escape the jet as CRs. The beam of the CR emission is therefore much larger than that of the neutrinos. Hence, UHECRs may be directly observed from AGN with greater viewing angles than sources producing point source neutrinos. }
\label{fig:schematic}
\end{minipage}
\end{figure*}

Protons are confined and accelerated (by shocks) near the base of the jet, and interact with ambient X-ray photon fields (see Fig. \ref{fig:schematic}). Following pion production, and subsequent decays, neutrinos and neutrons are produced and will escape the region. Whereas neutrinos escape unhindered, the neutrons will decay to CR protons, which would be observable. \citet{KT2008} use data from the  PAO to estimate the ultra high energy cosmic ray (UHECR) flux from Cen~A, and diffuse UHECR flux, above the threshold energy of PAO, $E_{\rm CR, th}=5.7\times 10^{19}{\rm eV}$. Due to the common production path of CR protons and neutrinos, and that the emission from Cen A is assumed representative for all sources, the UHECR flux is used to scale the neutrino flux.

There are two breaks in the UHECR proton spectrum, which are caused by the change in photo-pion production efficiency for the protons and neutrons. However, these breaks are close in energy, and the model therefore assumes a single break energy. This energy is determined through the $\gamma$-ray break energy, $E_{\gamma, {\rm br}} = 200~{\rm MeV}$ for Cen A \citep[see][and references therein]{KT2008}, such that $E_{\rm CR,br} = 3 \times 10^8 E_{\gamma,{\rm br}}$. The all-flavour neutrino flux from Cen A, $\Phi^{\rm Cen A}_{\nu_{\rm all}}$, using the UHECR proton flux $\Phi^{\rm Cen A}_{\rm p}$ as a normalization, can then be expressed by 
\begin{equation}\label{eqn:kt-cena-all}
\Phi^{\rm Cen A}_{\nu_{\rm all}} (E_\nu) 
	= \left[ 
   		\left( \frac{\xi_{\nu}}{\xi_{\rm n} \eta^2_{\nu \rm n}} \right) \min \left( \Xi , \Xi^{2} \right)   
   	  \right]
    	   \Phi^{\rm Cen A}_{\rm p}  \left(\frac{E_{\nu}}{\eta_{\nu \rm n}}\right) \  
\end{equation}
\citep[see][]{KT2008}, where $\xi_\nu$ and $\xi_{\rm n}$ are the fraction of proton energy that is converted to the  neutrino energy and the neutron energy respectively, and
\begin{equation}\label{eqn:kt-cena-all_psi}
	\Xi(E_\nu, E_{\rm CR, br}) = \frac{E_{\nu}}{\eta_{\nu \rm n} E_{\rm CR, br}} \ .
\end{equation}
Here the ratio of the average energy of neutrinos and neutrons is expressed as $\eta_{\nu \rm n} = \langle E_{\nu} \rangle/ \langle E_{\rm n} \rangle = 0.04$, and the fraction of the proton energy transferred to either neutrinos or neutrons in the initial interactions is given by $\xi_{\nu}/\xi_{\rm n}\approx 0.1/0.5 = 0.2$. The UHECR flux from Cen A above the threshold energy $E_{\rm CR,th}$ is $\Phi_{\rm p}^{\rm Cen A}(E_\nu)=5\times10^{-21} {\rm cm}^{-2} {\rm s}^{-1}$ \citep[CH;][]{KT2008}. The break in the neutrino spectrum can then be expressed in terms of the UHECR proton spectrum and the expression for ratio of average energies of neutrinos and neutrons, such that $E_{\nu, {\rm br}} \equiv \eta_{\nu \rm n} \, E_{\rm CR,br} = 4 \times 10^6~{\rm GeV}$. Due to the neutrino mixing from source to observed ratios, the muon neutrino spectrum is one third of the full neutrino spectrum, and the flux of muon neutrinos from Cen A is
\begin{equation}\label{eqn:kt-cena-mu}
\Phi^{\rm Cen A}_{\nu} = 
  A_{\nu}^{\rm [KT]} \left(\frac{E_{\nu}}{\rm GeV}\right)^{-\alpha_{\nu}} \left[ \min \left(1, \frac{E_{\nu}}{E_{\nu,{\rm br}}} \right) \right] \ ,
\end{equation}
with a proton power-law spectrum of index $\alpha_{\rm p} = 2.7$, and that of the neutrino spectrum $\alpha_\nu = 1.7$. The normalization factor, $A_{\nu}^{\rm [KT]}$ contains the scaling through the UHECR flux and the combination of energy contribution to the neutrinos from the initial particle interactions: 
\begin{eqnarray}\label{eqn:kt-anu}
A_{\nu}^{\rm [KT]} 	& = & \left[ \frac{(\alpha_{\rm p}-1)}{3} \left( \frac{\xi_{\nu} \eta_{\nu \rm n}^{\alpha_{\rm p}-2}}{\xi_{\rm n}} \right) 
   					 \frac{  E_{\rm CR,th}^{\alpha_{\rm p}-1}   }{E_{\nu,{\rm br}}}  \right]   \Phi_{\rm p}^{\rm Cen A}(E_{\rm th})    \nonumber \\			
				& \approx & 3 \times 10^{-11} {\rm GeV}^{-1} {\rm cm}^{-2}{\rm s}^{-1}  \ .
\end{eqnarray}
We note that a cut-off of the neutrino spectrum (due to it being limited by the maximum proton energy) is not included in \citet{KT2008}, as this will occur at the far end of the spectrum. We therefore assume a cut-off at an energy of $E_{\nu \rm , max} = 5\times10^{19}$ eV, at which the highest energy UHECR event is observed. 

To calculate the diffuse all-sky neutrino flux, \citet{KT2008} use two source models; one with AGN luminosity density of no evolution and another with strong evolution \citep{BT1998}, which lead to a diffuse flux $\sim 200-5000$ times larger than that of the Cen A flux. This implies a strong detection in either case when compared to the current experimental limit for neutrino detection in {\it IceCube} \citep[see Fig. \ref{fig:ktcena};][]{AAA2014b}. 

In this paper however, we use the Cen A neutrino spectrum as given in the \citet{KT2008} model, and convolve with AGN population densities derived from X-ray observations. Because Cen A is a typical source in this model, we use the ratio between the neutrino and X-ray luminosities as a reference, and scale the neutrino flux for an AGN of a given X-ray luminosity by this ratio:
\begin{equation}\label{eqn:agn_cena_ratio}
	\zeta^{\rm AGN} = \lambda\, \zeta^{\rm Cen A} \ ,
\end{equation}
using the simplest case, where the scaling factor $\lambda = 1$, and
\begin{equation}\label{eqn:cena_bratio}
	\zeta^{\rm Cen A} = \left(\frac{I_ { \nu}}{ I_{\rm X}}\right)^{\rm Cen A} \ .
\end{equation}
The X-ray photon intensity of Cen A is $I^{\rm Cen A}_ {\rm X} = L_{\rm X}/ 4{\mathrm \upi} D_{\rm L}^2 = 3.47 \times 10^{-10} {\rm erg \ cm}^{-2} \ {\rm s}^{-1}$. $D_{\rm L}$ is the luminosity distance, and for a local source it is the same as the measured proper distance. Thus for Cen A, $D_{\rm L} =  3.4~{\rm Mpc} =1.049 \times 10^{25} {\rm cm}$,
and a measured (2$--$10~keV) X-ray luminosity $L_{\rm X}^{\rm Cen A}=4.8 \times 10^{41} {\rm erg \ s}^{-1}$ \citep{EKW2004}. As Cen A is a local source, we calculate the single source spectrum at redshift $z=0$. For an AGN with X-ray luminosity $L^{\rm AGN}_{\rm X}$, we scale the spectrum for a single source with the Cen A brightness such that
\begin{equation}\label{eqn:cfb_linear_kt}
	\tilde{\Phi}^{\rm [KT,lin]} = \frac{L_{\rm X}^{\rm AGN}}{L_{\rm X}^{\rm Cen A}} \ ,
\end{equation}
giving the AGN fluxes generally,
\begin{equation}\label{eqn:sss_agn_cfb_linear}
	{\Phi}^{\rm AGN}_{\nu} = \Phi^{\rm Cen A}_{\nu} \tilde{\Phi}^{\rm [KT,lin]} \ .
\end{equation}
This linear scaling then reflects our expectation that a bright AGN produces a higher rate of neutrinos than a lower luminosity counterpart.

\subsection{The Becker \& Biermann (BB) model}
We compare the KT model predictions with a second model by \citet{BB2009}. Similar to the KT model, the BB model considers neutrinos of hadronic origin in AGN jets, and the initial seed protons are accelerated in shocks in the jet. However, in this model the peak of neutrino emission occurs further out in the jet, at the first stable shock, a distance of $r \sim 3000~r_{\rm g}$ (see Fig. \ref{fig:schematic}). For an additional comparison with the KT model, we have in Fig. \ref{fig:hillas_plot} added the location of Cen A if proton acceleration occurs at a location $r = 3000~r_{\rm g}$ (not accounting energy losses). 

After a comparison of photon optical depths, \citet{BB2009} find that the photon field that dominates the $p\gamma$ interactions is of synchrotron jet origin, with optical depth $\tau_{p\gamma} \sim 1$ for a bulk Lorentz factor of the jet $\Gamma \sim 10$. The frequency of the photon target field is therefore taken to be $f = 1~{\rm GHz}$. There are no breaks in the neutrino spectrum at these energies, as the break for the radio synchrotron photons occurs at much lower energies. 

Furthermore, 29 UHECR events observed by PAO appear to correlate to AGN in the super-galactic plane. The BB model therefore considers the UHECR, and hence neutrino, emission to originate in blazars and FR~I AGN. The neutrino spectrum is then normalized by the proton flux. 

The muon neutrino spectrum traces the proton spectrum, and has a cut-off at energies determined by the maximum energy of the energetic protons,
\begin{equation}\label{eqn:bb-mod-mu}
 \Phi^{\rm [BB]}_{\nu} = A_{\nu}^{\rm [BB]}  E_{\nu}^{-\alpha_{\nu}}  \exp{\left(-\frac{E_{\nu}}{E_{\rm max}}\right)} \ .
\end{equation}
The normalization $A_{\nu}^{\rm [BB]}$ is defined in terms of the redshift dependency factor, $\varphi_{\nu}/\varphi_{\rm CR}$, the ratio of the Lorentz factors, $\gamma_{\nu}$ and $\gamma_{\rm CR}$ of the neutrino and proton (CR) emission respectively; a measure of the optical depth in the source photon field, $\tau_{\rm p\gamma}$; the Auger threshold energy; and flux at energies larger than the minimum of the proton, $E_{\rm CR, min}=m_{\rm p} \approx \gamma_{\rm CR} \times {\rm GeV}$,
\begin{eqnarray}\label{eqn:bb-anu}
	A_{\nu}^{\rm [BB]} & = &   \frac{(\alpha_{\rm p}-1)}{12} 
	\left(  \frac{\varphi_{\nu}}{\varphi_{\rm CR}} \right)  \left(\frac{\gamma_{\nu}}{\gamma_{\rm CR}}\right)^{5-\alpha_{\rm p}} 
	     N(>E_{\rm CR, min})    \nonumber\\
					&& \hspace*{0.2cm} \times \  \tau_{\rm p\gamma}       \left(\frac{m_{\pi}}{4}\right)^{\alpha_{\rm p}-2}(E_{\rm PAO}^{\rm min})^{\alpha_{\rm p}-1}  \nonumber \\
					& =  &1.4 \times 10^{-10} \mathrm{GeV^{-1}cm^{-2}s^{-1}sr^{-1}} 
\end{eqnarray}
In the \citet{BB2009} model, neutrino emission needs to originate in blazar-type AGN to be detected, due to the beamed emission. A CR point source may on the other hand have a jet misaligned with our line of sight, as the emission cone of UHECRs is significantly larger than that of the neutrinos (see Fig. \ref{fig:schematic}). The redshift dependency ($\varphi_{\nu}/\varphi_{\rm CR}$) accounts for the difference in flux from neutrinos and CRs, based on the redshift evolution of their sources. Using radio luminosity functions (LFs) for FR~I type blazars
(BL~Lac sources)
and FR~I RGs, a ratio of the received emission of neutrinos and CRs, respectively, is estimated. The dependence is expressed as 
\begin{equation}
\varphi = \int^{z_{\rm{max}}}_{z_{\rm{min}}} \int^{L_{\rm{max}}}_{L_{\rm{min}}}  \mathrm{d}z\mathrm{d}L \frac{1}{4\upi D_{\rm{L}}^{2}} \frac{\mathrm{d}\Psi_{\rm{r}}}{\mathrm{d}L} \frac{\mathrm{d}V_{\rm{c}}}{\mathrm{d}z} \ ,
\end{equation}
with the radio LF $\mathrm{d\Psi_{r}}/\mathrm{d}L$ and the comoving volume element $\mathrm{d}V_{\rm{c}}/\mathrm{d}z$. The factor $1/4\upi D_{\rm{L}}^{2}$ takes into account the decrease of flux suffered for emission from sources at luminosity distance $D_{\rm{L}}$. To find the ratio between the redshift dependency of the emission, \citet{BB2009} use the flat spectrum radio source LF from \citet{DP1990}, and the FR~I radio LF given in \citet{WRB2001}, and estimate a value of $\varphi_{\nu}/\varphi_{\rm CR} \approx 0.1$.

We will on the other hand use the source densities, as we did for the KT model, based on X-ray observations, and thus XLFs for the AGN populations. To scale the neutrino spectra, we use the disc-jet symbiosis model \citep{FB1995} that relates the radio power of the jet to the disc luminosity, 
\begin{equation}
\frac{L_{\rm jet}}{L_{\rm disc}} = \kappa_{\rm d-j} \ ,
\end{equation}
adopting disc-jet parameter $\kappa_{\rm d-j} = 0.15$ from \citet{BBR2005}. We use two estimates of the scaling between the radio and total power in the jet, giving two luminosity scaling models. In the first case, we use the synchrotron to total jet luminosity relation \citep{CMN2010}
\begin{equation}
	L_{\rm jet} \approx 5.8  \times 10^{43} \left(\frac{L_{\rm synch}}{10^{40} \ {\rm erg~s}^{-1}   } \right)^{0.7} {\rm erg~s}^{-1} \ .
\end{equation}
This gives a disc-jet scaling in terms of the synchrotron luminosity,
\begin{equation}
	L_{\rm synch}^{\rm [BB1]} \approx 2.0  \times 10^{-24} (L_{\rm disc})^{1.43} \ {\rm erg~s}^{-1} \ .
\end{equation}
The second scaling model is adopted from \citet{BBR2005}, and relates the synchrotron luminosity to the disc luminosity by
\begin{equation}
	L_{\rm synch}^{\rm [BB2]} \approx 3.3  \times 10^{-15} (L_{\rm disc})^{1.27} \ {\rm erg~s}^{-1} \ ,
\end{equation}
following \citet{FB1995} and \citet{FMB1995}. The scaling model can be introduced to the neutrino energy calculations as the synchrotron luminosity of the AGN is proportional to the optical depth for $p\gamma$ interactions, and the optical depth is also proportional to the normalization factor for the neutrino spectrum \citep{BB2009}. The optical depth then gives the scaling
\begin{equation}
\tilde{\Phi}^{\rm [BB_{1}]} = \left( \frac{L_{\rm X}}{3.87\times 10^{44} \ {\rm erg \ s^{-1}}}\right)^{1.43}  \ ,
\end{equation}
and
\begin{equation}
\tilde{\Phi}^{\rm [BB_{2}]} = \left( \frac{L_{\rm X}}{1.04 \times 10^{43} \ {\rm erg \ s^{-1}}}\right)^{1.27}  \ ,
\end{equation}
giving the AGN flux 
\begin{equation}
	{\Phi}^{\rm AGN}_{\nu} = \Phi^{\rm [BB]}_{\nu} \tilde{\Phi}^{\rm [BB_{i}]} \ .
\end{equation}
We use a jet Lorentz factor, $\Gamma_{\rm jet} = 10$ and a jet half opening angle of $\omega_{1/2} = \Gamma^{-1}_{\rm jet} = 0.1~{\rm rad}$. The fraction of the luminosity of a knot to the total synchrotron luminosity is $\epsilon = 0.1$, and the neutrino production occurs at a distance of 3000 gravitational radii.

\section{Evolving AGN populations}\label{sec:agn_populations}
We consider AGN number evolutions from two X-ray surveys, selecting X-ray emitting AGN \citep[thus across the entire population, regardless of radio-loudness;][]{SGB2008}, and blazars \citep{ACS2009}. The two give us sets of widely different samples of AGN, both varying in luminosity and viewing angle, and where the former contains both RL and RQ AGN, and the latter contains only a fraction of the RL population. This is accounted for when we derive the total number of AGN in the Universe, using the prescription of the XLF given in these works.

Neutrino emission may be directly scaled with $\gamma$-rays originating from pion decays \citep[e.g.][]{HZ1997,AD2001,NR2009}. However, only an upper limit can be set on the neutrino emission, as some fraction of the emitted $\gamma$-rays would originate from the upscattering of e.g. internal synchrotron photons. Moreover, the $\gamma$-ray emission originates in blazar sources, with highly beamed luminosities, and are therefore suitable for a point source study. In this paper we aim to explore the neutrino emission from all jetted AGN classes, hence require a survey which is insensitive to orientation effects and obscuration of the jet component. As X-rays trace the accretion power of the AGN system, a survey in this waveband fits the purpose. 

Furthermore, by choosing X-ray surveys as our AGN study, there is no need for complementary observations at different wavebands, as X-ray emission implies accretion on to a SMBH at the AGN centre. We can therefore be confident that all X-ray luminous objects observed with a X-ray luminosity above $L_{\rm X}=10^{42}~{\rm erg~s}^{-1}$ are AGN \citep[see e.g.][]{TUC2004}.

\citet{SGB2008} measure the hard (2-8~keV) XLF of AGN up to $z \sim 5$. The sample consists of 682 AGN in total, with 31 found at redshifts $z>3$. They use the {\it Chandra} multi-wavelength project to detect high redshift luminous AGN ($L_{\rm X}>10^{44}~{\rm erg~s}^{-1}$), and the {\it Chandra} deep field to cover the lower luminosity range. \citet{ACS2009} have used 3 yrs of data from the {\it Swift}/BAT survey to select a complete sample of X-ray blazars to determine the evolution of blazars in the 15-55~keV band. The sample consists of 26 FSRQs and 12 BL~Lac objects in a redshift range of $0.03<z<4.0$. 

Both AGN population surveys show that the source density evolution of AGN depends on luminosity and epoch. The number density can therefore be derived using the XLF, assuming that the luminosity distribution of the neutrino sources are complete and as inferred by \citet{SGB2008} and \citet{ACS2009}, and can be extrapolated to redshifts up to $z=10$.

\subsection{The X-ray luminosity function}\label{ssec:xlf}
%
\begin{table*}
\begin{minipage}{16.5cm}
\caption{Fitted XLF model parameters. Summary of the fitted parameters for each best-fitting XLF model, as given in the relevant papers: model with first letter S refers to model from \citet{SGB2008}; models with first letter A refers to the models given in \citet{ACS2009}. LDDE: luminosity dependent density evolution; MPLE: modified pure luminosity evolution, and PLE refers to the pure luminosity evolution. In brackets are the source population modelled, such that Type I/IIs are described by an LDDE model \citep[model B in][]{SGB2008}, blazars and FSRQs by MPLE models \citep[best-fitting models 7 and 10, respectively;][]{ACS2009}, and BL Lacs by the PLE (best-fitting model 11).}
\label{table:xlf_param}
\begin{tabular}{lccclccccccc}
\hline 
\multirow{2}{*}{Model} 
	& \multicolumn{4}{l}{$z=0$ Parameters} 
	& \multicolumn{5}{l}{evolution parameters} \\
	& $(A, \log A)$ & $(L_{\ast}, \log L_{\ast})$ & $(\Upsilon_1)$ & $(\Upsilon_2, \tilde{\Upsilon}_{2})$ 
	& $(\upsilon_{1}, \tilde{\upsilon}_{1})$ & $(\upsilon_{2},  \tilde{\upsilon}_{2})$ & $(z_{\rm c})$ & $(\log L_{\rm c})$ & $(\alpha)$  \\
\hline
SLDDE (Type I/II)& $ -6.077^{a} $ & 44.33 & 2.15 & 1.10 & 4.00 & $-1.5$ & 1.9 & 44.6 & 0.317 \\
AMPLE (Blazar) & $1.379 \times 10^{-7}$ & $1.81^{b}$ & $-0.87$ & 2.73 & 3.45$^{e}$ & $-0.25^{e}$ & -- & -- & --  \\
AMPLE (FSRQ) & $0.175 \times 10^{-7}$ & $2.42^{b}$ & $<-50.0^{c}$ & 2.49 & 3.67$^{e}$ & $-0.30^{e}$ & -- & -- & --  \\
APLE (BL Lac) & $0.830 \times 10^{-7}$ & $1.0^{b}$ & -- & 2.61$^{d}$ & $-0.79^{e}$ & -- & -- & -- & --  \\
\hline
\end{tabular}
\medskip
{\it Notes.} $^{a}$The value represents the logged normalisation constant, $\log A$, as is given in \citet{SGB2008}. %
$^{b}$The value represents the unlogged value of the break luminosity, $L_{\ast}$, where the luminosities are all normalised to $L_{44}=10^{44}~{\rm erg~s}^{-1}$. %
$^{c}$In these calculations we used $\gamma_1 = -50.0$ \citep[see][]{ACS2009}. %
$^{d}$The BL Lac XLF model uses the single power law expression (Equation \ref{eqn:pd_xlf_1}), with index $\tilde{\Upsilon_{2}}$. %
$^{e}$The blazar, FSRQ and BL Lac XLFs assume an evolution defined by the indices $\tilde{\upsilon}_{1}$ and $\tilde{\upsilon}_{2}$.
\end{minipage}
\end{table*}
The differential XLF of a population is a measure of the number of objects per comoving volume and unit luminosity interval, as a function of X-ray luminosity and redshift,
\begin{equation}\label{eqn:xlf-def}
	\frac{\mathrm{d} \Psi (L_{\rm X}, z)}{\mathrm{d} \log L_{\rm X}} =  \frac{\mathrm{d}^2 N(L_{\rm X},z)}{\mathrm{d} V_{\rm c} \mathrm{d} \log L_{\rm X}} \ .
\end{equation}
The present-day XLF can be expressed as a simple power-law \citep{ACS2009},
\begin{equation}\label{eqn:pd_xlf_1}
\frac{\mathrm{d}\Psi (L_{\rm X}, z=0)}{\mathrm{d} \log L_{\rm X}} = A \ \ln(10) \left(\frac{L_{\rm X}}{L_{\ast}}\right)^{1-{\tilde \Upsilon}_2} \ ;
\end{equation}
however, observationally there is a break, and with a high enough source count, this break can be seen. A double power-law \citep[e.g.][]{UAOM2003} can fit the observational data, with the faint and bright end slopes dictated by  $\Upsilon_1$ and $\Upsilon_2$, respectively, for luminosities below and above the break luminosity $L_{\ast}$, such that 
\begin{equation}\label{eqn:pd_xlf_2}
	\frac{\mathrm{d} \Psi (L_{\rm X}, z=0)}{\mathrm{d} \log L_{\rm X}} 
	 =  A \left[ \left(\frac{L_{\rm X}}{L_{\ast}} \right)^{\Upsilon_1}+ \left(\frac{L_{\rm X}}{L_\ast} \right)^{\Upsilon_2} \right]^{-1} \ .
\end{equation}

The parameters in the XLFs are determined through maximum likelihood routines, using the \texttt{MINUIT} minimization package \citep[see][for details]{SGB2008,ACS2009}, and here we use the best-fitting values as given in the respective papers (see Table \ref{table:xlf_param}). We also note that 
\begin{equation}\label{eqn:pd_xlf_log}
	\frac{\mathrm{d} \Psi (L_{\rm X}, z=0)}{\mathrm{d} \log L_{\rm X}} =  A \ln (10)~L_{\rm X} \frac{\mathrm{d} \Psi (L_{\rm X}, z=0)}{\mathrm{d} L_{\rm X}}  \ .
\end{equation}
The evolution of the XLF depends on the chosen model that fits the observations best. The base models are the pure luminosity evolution (PLE) and the pure density evolution (PDE), however these are not found to represent the observational data well. Therefore, modified versions of these models are used, either extending the form of the luminosity or density evolution, or formulating a combination of the two \citep[see e.g.][]{UAOM2003,ANL2010}. 

The blazar population is found to be best described in terms of a modified pure luminosity evolution (MPLE) model \citep{ACS2009} on a double power-law present-day XLF (Equation \ref{eqn:pd_xlf_2}), where the evolution factor is a power law with two free parameters, $\upsilon_{1}$ and $\upsilon_{2}$, giving a general behaviour with respect to redshift, of a form first given in \citet{WPS2008},
\begin{equation}\label{eqn:ple}
\frac{\mathrm{d} \Psi(L_{\rm X}, z)}{\mathrm{d} \log L_{\rm X}} = \frac{\mathrm{d} \Psi [L_{\rm X}/e(z), 0]}{\mathrm{d} \log L_{\rm X}} \ ,
\end{equation}
and
\begin{equation}
e(z)= (1+z)^{{\tilde\upsilon}_{1}+{\tilde \upsilon}_{2} z} \ .
\end{equation}
The fitted parameters are summarized in Table \ref{table:xlf_param}.

We also take a closer look at the best-fitting XLFs of the subclasses of the blazars, namely the BL Lacs and FSRQs. Whereas the FSRQs are modelled similarly to the full blazar sample, the BL Lacs are too few in number, so we use the best-fitting simple power-law XLF of \citet{ACS2009} (Equation \ref{eqn:pd_xlf_1}), with a simple evolution factor, $e(z)= (1+z)^{{\tilde \upsilon}_{1}}$ (see Table \ref{table:xlf_param}).

\citet{SGB2008} determined that the best-fitting XLF for their sample is the luminosity dependent density evolution (LDDE) model, for which the evolution factor, $e(z,L_{\rm X})$ is a function of both redshift and luminosity. It is convolved with the double power law present-day XLF (Equation \ref{eqn:pd_xlf_2}) to determine the population density evolution as follows:
\begin{equation}
\frac{\mathrm{d} \Psi(L_{\rm X},z)}{\mathrm{d} \log L_{\rm X}} = \frac{\mathrm{d} \Psi (L_{\rm X}, z=0)}{\mathrm{d} \log L_{\rm X}}e(z, L_{\rm X}) \ .
\end{equation}
The evolution factor is defined in terms of a luminosity dependent redshift cut-off $z_{\ast}$, which is further determined 
by a power law of $L_{\rm X}$,
\begin{equation}
	e(z, L_{\rm X})=
		\begin{dcases}
			(1+z)^{\upsilon_{1}} & [z< z_{\ast}(L_{\rm X})] \\
				e(z_{\ast}(L_{\rm X})) \left[ \frac{1+z}{1+z_{\ast}(L_{\rm X})}\right]^{\upsilon_{2}} & [z \geq z_{\ast}(L_{\rm X})]
		\end{dcases} \ ,
\end{equation}
and
\begin{equation}\label{cutoff_zc}
	z_{\ast} (L_{\rm X})=
		\begin{dcases}
			z_{\rm c} \left[\frac{L_{\rm X}}{L_{\rm c}}\right]^{\alpha} & (L_{\rm X}< L_{\rm c}) \\
				z_{\rm c} & (L_{\rm X} \geq L_{\rm c})
		\end{dcases} \ .
\end{equation}
Another five parameters are therefore introduced when evolving the XLF; to determine the redshift cut-off, the characteristic luminosity $L_{\rm c}$, the cut-off redshift $z_{\rm c}$ and the strength of the redshift cut-off dependence $\alpha$. In addition, the evolution rates prior to and beyond the redshift cut-off $z_{\ast}$ are given by $\upsilon_{1}$ and $\upsilon_{2}$, respectively. 

We assume the XLF at lower redshifts can be extrapolated to describe the high-redshift evolution, and as such we span the AGN evolution from redshifts $0<z<10$. We set the upper redshift to $z=10$, however note that the oldest quasar is found at a redshift $z \approx 7$ thus far \citep{MWV2011}. The lower and upper luminosity bounds on the AGN populations are carefully determined, particularly for the FSRQ population, as the faint end of the FSRQ XLF collapses towards higher luminosities dependent on the redshift bin, seen in Fig. \ref{fig:agn_plots}. This will be discussed further in the next section (\ref{ssec:zl_distributions}).

We carry out our calculations assuming the distribution of luminosities obtained from these XLFs is a good representation of AGN sources -- though we note the possibility of missing a low-luminosity AGN contribution in the surveys, especially at high redshifts. This is mentioned in \citet{ACS2009}, as BAT is not sensitive to low-luminosity and low-redshift sources. The faint end might be under representative of the real population, as indicated by their results and earlier radio-selected surveys of blazars. 
 
We use the XLFs to calculate the number densities, over a range of luminosities and redshifts. This enables us to study and compare the neutrino contribution predicted from AGN of low and high luminosities, and also from low to high redshifts. The comoving volume for a flat, matter dominated cosmology, is measured \citep{P2007} as 
\begin{equation}
\frac{\mathrm{d}V_{\rm c}}{\mathrm{d}z} =16 \upi \left(\frac{c}{H_0} \right)^3 \frac{(\Omega z+(\Omega-2)[\sqrt{1+\Omega z}-1])^2}{\Omega^4 (1+z)^3\sqrt{1+\Omega z}} \ ,
\end{equation}
and we use the cosmological prescriptions given in the relevant papers to maintain consistency of each population ($H_0 = 70~{\rm km~s}^{-1} \ {\rm Mpc}^{-1}$, $\Omega_{\Lambda}=0.7$, $\Omega_{\rm M}=0.3$). The X-ray luminosities are normalized to $L_{44}=10^{44}\ {\rm erg~s}^{-1}$ in our calculations. 

As an aside, it is worth emphasizing that our calculations are conservatively based only on the well-understood AGN populations. A surprising new radio AGN class `FR~0' was recognized recently: their radio cores resemble FR~I cores, lacking extended
radio emission
\citep{BC2009,G2011},
and they may outnumber FR~I sources by $\sim3$ -- $\sim100$ times \citep{SEM2014,BCG2015}.
Why the FR~0 cores fail to drill their jets farther out is unclear --
   perhaps due to youth, intermittency, interstellar medium obstruction
	\citep[like GPS/CSS sources; e.g.][]{OD1998,SBS2005},
   or intrinsically low $\Gamma$ or slow SMBH spin \citep{BCG2015}.
Whatever the reason,
   if FR~0 cores turn out to be as $\nu$-bright as FR~I cores,
   then their addition would {\em strengthen} our constraints on overall AGN neutrino production.
The limits also tighten in a similar manner if,
   for instance,
   our $L_\mathrm{X}$ cut-off
   has underestimated a significant contribution from
   lower luminosity FR~I RGs \citep[e.g.][]{balmaverde2006,hardcastle2009,capetti2015}.

\subsection{AGN number density distribution}\label{ssec:zl_distributions}
We compute the evolutionary tracks and luminosity distributions over several cosmological epochs for all four AGN subpopulations, integrating the XLFs with respect to luminosity and redshift, respectively. To obtain estimates for the full AGN population, we scale the XLF by a correction factor $\Theta_{\rm CF}$ to obtain the number of all AGN within our redshift range. For the RG sample we account for those obscured or too faint following \citet{ZMI2011}, and assume that observed sources are 10\% of the total population. However, the \citet{SGB2008} survey collects both RL and RQ sources, so we assume the RG population accounts for 10\% of all X-ray detected AGN \citep{UP1995}. The correction factor for RGs is therefore $\Theta_{\rm CF} = 1$. 
%
\begin{table*}
\begin{minipage}{13.3cm}
\caption{AGN space densities. The space densities calculated for the various AGN populations considered in a redshift range $0<z<10$ are shown. The luminosity ranges assumed for each population are summarised together with the corresponding unbeamed luminsity. For the blazar population we determined this intrinsic luminosity to avoid contamination from other beamed luminous sources at high redshift (see the text).}
\label{tab:lum_spacedensity}
\begin{tabular}{lcccc}
\hline
							& RG 					& Blazars				& FSRQ (FR~II)			& BL Lac (FR~I) \\
\hline
${L_{\rm X}~[{\rm erg~s^{-1}}]}$:	& $10^{42}$-$10^{47}$		& $10^{44}$-$10^{48.5}$		& $10^{46}$-$10^{48.5}$ 		& $10^{44.5}$-$10^{48.5}$  \\
$\mathcal{L}_{\rm X}~[{\rm erg~s^{-1}}]$:	& $-$			& $10^{40}$-$10^{44.5}$		& $10^{43.8}$-$10^{48.5}$ 	& $10^{40}$-$10^{44}$	  \\
\bf{Space density [Mpc$^{-3}$]}:	& $\sim 1.36 \times 10^{-4}$	& $\sim 2.26 \times 10^{-4}$	& $\sim 6.84 \times 10^{-6}$	& $\sim 5.79 \times 10^{-4}$ \\
\hline
\end{tabular}
\end{minipage}
\end{table*}
%
In the case of blazars we correct for misaligned sources, obtaining the correction factor as the ratio of the solid angle of a full sphere to the solid angle of the jet projection on to this sphere. The viewing angle is $\omega_o = \omega_{1/2}$, so that a jet with an opening angle of $2\omega_{1/2}$ will be not be in our line of sight if the viewing angle is larger than the half opening angle. Assuming a modest bulk Lorentz factor $\mathrm{\Gamma = 10}$, which relates to the half opening angle by $\omega_{1/2}=\Gamma^{-1} \approx 5\fdg7$, these misaligned sources imply that
\begin{align}
\Theta_{\rm CF} &= \frac{4\upi}{\Omega} = \frac{4\upi}{2\upi(1-\cos(\omega_{1/2}))} \nonumber \\
			&\approx \frac{4\upi}{2\upi(\omega_{1/2}^2/2)} = 4\Gamma^2 \ .
\end{align}
This gives a correction factor of 400, which agrees with estimates of a few hundred, or $2\Gamma^2$ \citep{ACS2009, GDV2010, VHG2011}. We calculate the AGN number density evolution over cosmological epochs by integrating the XLF into bins of X-ray luminosity to give the redshift distribution, such that
\begin{equation}
\frac{\mathrm{d}N(z)}{\mathrm{d}z} = \Theta_{\rm CF} \int^{\log L_2}_{\log L_1} \frac{\mathrm{d} \Psi (L_{\rm X},z)}{\mathrm{d} \log L_{\rm X}} \frac{\mathrm{d}V_{\rm c}}{\mathrm{d}z} \mathrm{d}\log L_{\rm X} \ .
\end{equation}
The luminosity dependence of the AGN population found in the XLF prescriptions motivates a closer look at the luminosity distribution of AGN in bins of redshift. We assess how the dominant luminosity class varies with redshift by integrating the XLFs over several cosmological epochs in the range of X-ray luminosity adopted in our calculations, which gives
\begin{equation}
\frac{\mathrm{d}N(\log L_{\rm X})}{\mathrm{d} \log L_{\rm X}} = \Theta_{\rm CF} \int^{z_2}_{z_1} \frac{\mathrm{d} \Psi (L_{\rm X},z)}{\mathrm{d} \log L_{\rm X}} \frac{\mathrm{d}V_{\rm c}}{\mathrm{d}z} \mathrm{d}z \ .
\end{equation}

We use the appropriate luminosity range, with each bin spanning an equal size for a consistent comparison. We choose the upper and lower bounds by evaluating the maximum luminosity of an AGN, according to the Eddington luminosity of a given SMBH mass. For AGN, we assume an upper mass of $M_{\bullet} \sim 10^{9} {\rm M}_{\Sun}$, and we find that the maximum luminosity should be about $10^{47}\ {\rm erg~s}^{-1}$. Thus for the RL derived AGN population, our range follows that of \citet{SGB2008}, spanning six orders of magnitude. 

In the case of blazars, we need to account for the beaming of these objects, as the quoted X-ray luminosities given in \citet{ACS2009} are referring to observed luminosities. The jets of observed blazars are beamed in our direction, hence the X-ray luminosities we record for the sources are greatly enhanced by this phenomenon. As the blazar surveys probe the deep past of the Universe, the estimation of the population size is based on these luminosities, and we therefore make a cut at an intrinsic luminosity of $10^{40}~{\rm erg~s}^{-1}$. We will only take those above this luminosity to be actual blazar observations, as we otherwise may confuse some of those that fall below this luminosity with e.g. X-ray binaries and other luminous objects that could also be observed at these redshifts \citep[e.g.][]{SGTW2004,FS2011}. These sources may also show beamed luminosities comparable to the fainter blazars, but whose intrinsic luminosity generally is found at around $10^{38}~{\rm erg~s}^{-1}$.

We calculate the intrinsic X-ray luminosity following \citet{US1984}, using the relation between observed $L_{\rm X}$ and emitted luminosity $\mathcal{L}_{\rm X}$ for a relativistic jet,
\begin{equation}\label{eqn:beaming_l}
L_{\rm X} = \delta^{\varrho} \mathcal{L}_{\rm X} \ ,
\end{equation}
where $\delta = [\Gamma (1-\beta \cos (\theta))]^{-1}$ is the jet Doppler factor, $\beta$ is the velocity in terms of the speed of light, and the Lorentz factor $\Gamma = [1-\beta^2]^{-1/2}$, and the viewing angle $\omega_{o} = \Gamma^{-1}$. The exponent $\varrho$ gives the enhancement of the luminosity, and for a blazar type in which only one jet is seen, $\varrho = 3 + \alpha$, where $\alpha$ is the spectral index. This exponent is due to relativistic aberration, whereby the emission is beamed forward due to the relativistic motions of the jet; contraction of the time interval, thus we observe more photons per unit time; and the blueshifting of photons, as there are a factor $\delta^\alpha$ more photons at the observed frequency than at the emitted frequency. It is found that the observed and intrinsic LFs have the same slope for high luminosities, however the observed LF will flatten towards lower luminosities because it is sensitive to the lower cut-off and steepness of the Lorentz factor distribution \citep{L2003}.

\begin{figure*}
\begin{minipage}{17cm}
\centering
\resizebox{0.8\hsize}{!}{\includegraphics[angle=0]{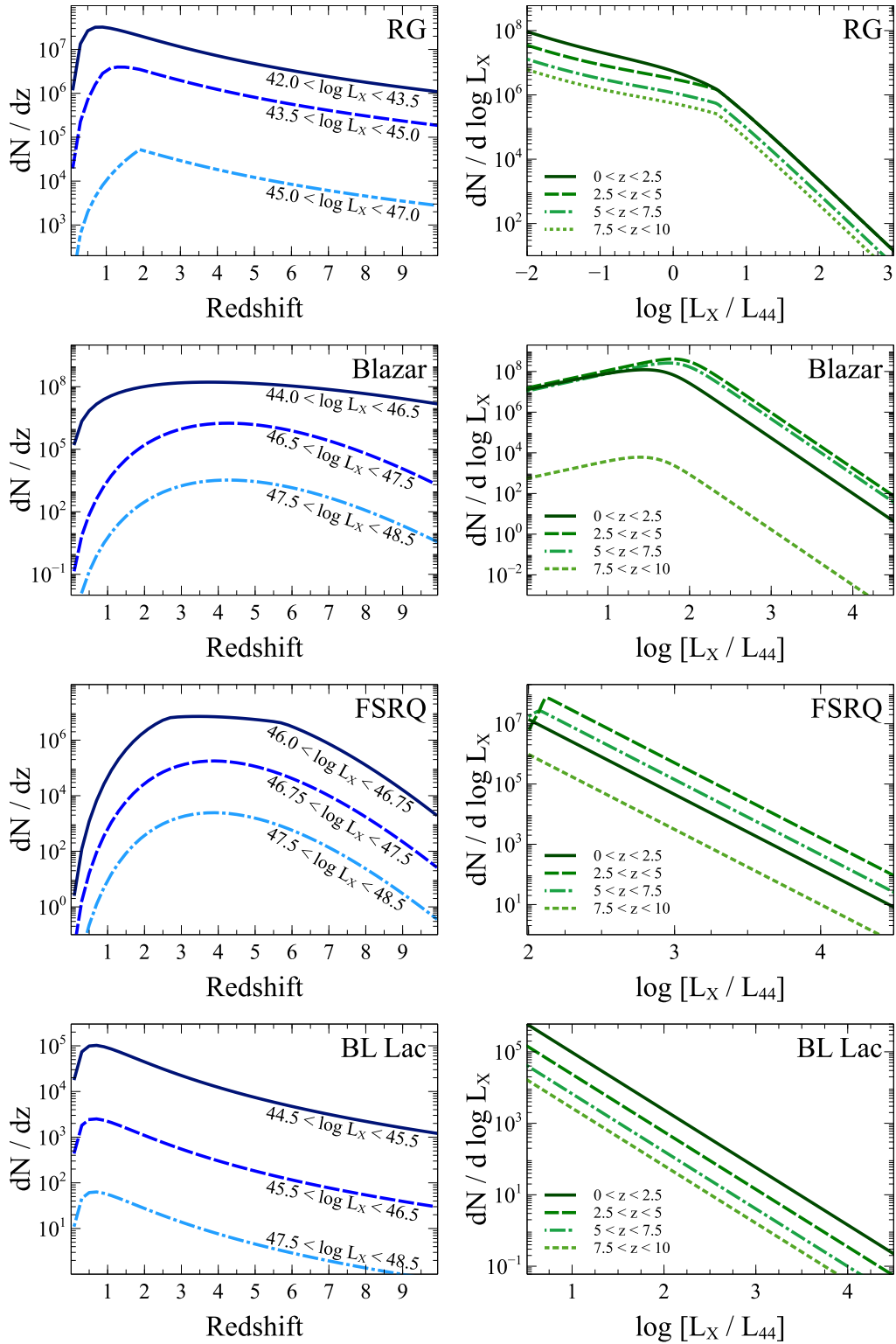}} 
\caption{\textbf{AGN redshift and luminosity distributions.} The panels on the left are the redshift distributions of RGs, blazars, FSRQs and BL Lacs (top to bottom). The panels on the right give the corresponding luminosity distributions. The overall trend is that the brighter AGN peak at earlier times, whereas the low-luminosity AGN are abundant at lower redshifts. Furthermore, the bright quasars are seen to dominate at higher redshifts, suggesting the density evolution of the brightest AGN was exceeded by fainter AGN at a redshift of $ z\sim 1$. See the text for details.}
\label{fig:agn_plots}
\end{minipage}
\end{figure*}

We assume a representative value for the spectral index of sources in a given AGN population. We use, for the subsamples BL Lacs, $\alpha = 1.5$; for FSRQs, $\alpha = 0.6$; and the total blazar sample we use $\alpha = 1.0$ \citep[see fig. 2][]{ACS2009}. This means that the lowest luminosity bound for the full blazar-derived population and the FSRQ-derived population is $10^{44} \ {\rm erg~s}^{-1}$, and for BL Lacs a little higher, at $10^{44.5} \ {\rm erg~s}^{-1}$. A further consideration is in order for the FSRQ-derived population, because the XLF for these sources collapses at lower luminosities, as seen in the FSRQ luminosity distribution in Fig. \ref{fig:agn_plots}. We therefore make a lower cut for this population at $10^{46} \ {\rm erg \ s}^{-1}$. The assumed X-ray luminosity ranges for the AGN populations are summarized in Table \ref{tab:lum_spacedensity}.
We find that the cuts we have made do not affect the total estimated numbers significantly. 

The large-scale space density (between $0<z<10$) is derived for the various populations, given the comoving volume contained within a redshift of $z=10$ is $V_{\rm c} \approx 3.5 \times 10^{12}~{\rm Mpc^{3}}$ \citep{W2006}. These are summarized in Table \ref{tab:lum_spacedensity}, and agree with local AGN estimates from UHECR observations \citep[see e.g.][and references therein]{TIY2012}.
%
%
We find that the higher luminosity AGN are preferentially found at higher redshifts. The blazar population is dominated by FSRQs at high luminosities ($L_{\rm X}> 10^{46}~{\rm erg~s}^{-1}$), and peaks here, with the brightest objects most numerous in the redshift bin between 4 and 5. This peak moves towards lower luminosities in more recent epochs. 

The RG population on the other hand, has a higher number density in the redshifts between 1 and 3, and similarly to the blazars, its brightest objects are found to peak in this range. In the local Universe the fainter RGs dominate, and beyond the peak (i.e. $z>2.5$), the source density is declining. The BL Lacs are lower luminosity objects, that are declining in numbers with higher luminosity, and the population dominates in the near Universe, in redshifts around $0<z<2$.

Furthermore, at earlier cosmological epochs, we find the higher luminosity AGN dominate. There is therefore a deficiency of bright AGN in the local Universe. Thus, it is suggested that the most luminous blazars formed early in the Universe, followed by a quick decrease in density \citep{ACS2009}. Though this implies that black holes formed early on in the Universe, and that early Universe conditions were favourable to the formation of very luminous AGN, the observational limitations at higher redshifts restricts the study of low-luminosity AGN at the same distances. 

To check our results from these two AGN population estimates, we sum the contributions from either the luminosity bins or redshift bins, and find that the sums agree, giving the space densities as quoted in Table \ref{tab:lum_spacedensity}. 

In view of neutrino output, we would thus expect a high production rate from bright quasars from the epoch of $z\sim4$, and bright AGN at $z\sim2$. If neutrino emission scales with the X-ray luminosity, these sources should then produce a higher rate of neutrinos than their lower luminosity counterparts. However, the fainter sources are more numerous, hence, despite a lower rate of neutrino production, the sheer number of these sources would imply a considerable contribution at more local redshifts.

\section{High-energy neutrino spectra}\label{sec:nu_spectra}
We make use of the neutrino production models described in Section \ref{sec:nu_models}, combined with the number distribution of neutrino sources from the AGN evolution models in Section \ref{sec:agn_populations}, to calculate the expected high-energy spectra, assuming that AGN are indeed the dominant high-energy neutrino machines (and therefore ignoring other possible sources for simplicity). 

We note the distributions of AGN both in luminosity and redshift, and find that by applying a luminosity scaling, we would expect to find the highest contribution of neutrinos from low-luminosity AGN at later times, as these sources are the most numerous in this epoch. Due to high-luminosity AGN dominating at earlier times (i.e. at redshifts $z > 4$), these should be prominent contributors, as their neutrino production rate should be considerably higher than in their low-luminosity counterparts. 

The luminosity scaling model is a simple relation between the neutrino luminosity and the X-ray luminosity of the source. This assumes the X-rays originate in the accretion disc, as commonly attributed. Observationally, the radiative and kinetic jet power correlates linearly with the disc luminosity \citep{ghisellini2014}. The KT neutrino output is scaled by a ratio of the AGN X-ray luminosity to that of the representative source, Cen A. The other scaling model, applied to the BB neutrino spectra, links the disc X-ray emission to the jet synchrotron emission, which ties the neutrino emission to the disc X-ray luminosity indirectly (see Section \ref{sec:nu_models}). Thus, a more powerful AGN would be brighter in X-rays. Similarly, a more powerful AGN will have greater potential to accelerate particles in its jet, and contributing to a higher rate of interactions. This again leads to an expected higher rate of neutrino production in these jets. We can therefore link the X-ray luminosity of the AGN (i.e. a direct measure of the accretion power of the AGN disc) to the neutrino luminosity (a consequence of the available energy in the AGN jet to accelerate and accommodate particle interactions of energies related to the AGN power). 

We produce a single source neutrino spectrum, following the published models outlined in Section \ref{sec:nu_models}, and convolve this with the AGN data to obtain emission from the entire populations. We scale with luminosity to reflect the influence that the source power has on the rate of neutrino production. We also correct for cosmological effects. The neutrino flux spectrum is required in terms of observable emission on Earth, and we calculate our spectra in the source frame. Hence, we shift our single source to different cosmological epochs, such that the emitted energy at source, $E_{\rm int}$, is related to the energy as we would observe it on Earth, $E_{\rm obs}$ through 
\begin{equation}\label{eqn:energy_shift}
E_{\rm int}=E_{\rm obs} (1+z) \ . 
\end{equation}
We carry out the spectral calculations, and relate the intensity received on Earth, $I_{\rm obs}$, to the intensity calculated at source, $I_{\rm int}$, 
\begin{equation}\label{eqn:intensity_shift}
I_{\rm obs}=I_{\rm int}(1+z)^{-4} \ ,
\end{equation}
to obtain the neutrino flux expected to be observed on Earth, taking into account cosmological effects such as redshift distortions. For a typical source, the intensity of the neutrino emission drops significantly with increasing redshifts, and a break in the spectrum will move to lower energies (as demonstrated in the KT model calculations). We obtain the neutrino spectra produced in AGN populations from various cosmological epochs by 
\begin{equation}\label{eqn:nz_tot}
E_\nu^2 \Phi^{\rm [model]}_{\nu} =  E_\nu^2 \Phi^{\rm AGN}_{\nu} \frac{\mathrm{d}N(z)}{\mathrm{d}z}~ \Delta z \ ,
\end{equation}
and produced by AGN of certain luminosities by
\begin{equation}\label{eqn:nl_tot}
E_\nu^2 \Phi^{\rm [model]}_{\nu} =  E_\nu^2 \Phi^{\rm AGN}_{\nu} \frac{\mathrm{d}N(\log L_{\rm X})}{\mathrm{d}\log L_{\rm X}}  \Delta [\log L_{\rm X}] \ .
\end{equation}
The sum of the binned contributions in each case gives the total diffuse emission as we would observe it. We measure the spectra against the experimental flux limit set by {\it IceCube}, 
\begin{equation}
	E^{2}_{\nu} \Phi_{\nu} \ {\rm [GeV \ cm^{-2} \ s^{-1} \ sr^{-1}]} \leq 1.44 \times 10^{-8} \ ,
\end{equation}
in the energy range $3.45\times 10^{4} < E_{\nu} [{\rm GeV}] < 3.66 \times 10^{7}$, determined with 1 yr of data \citep{AAA2014b}. This is an estimate of the minimum neutrino flux required for detection, and therefore gives an upper bound on the neutrino flux, as {\it IceCube} has detected only a few tens of events so far \citep[][]{AAA2013, IC2013}. 

\subsection{Resultant neutrino spectra}
We present a representation of observable neutrino emissions originating in various cosmological epochs, or from a range of source luminosities. 

\subsubsection{KT model spectra}
\begin{figure}
\centering
\resizebox{1\hsize}{!}{\includegraphics[angle=0]{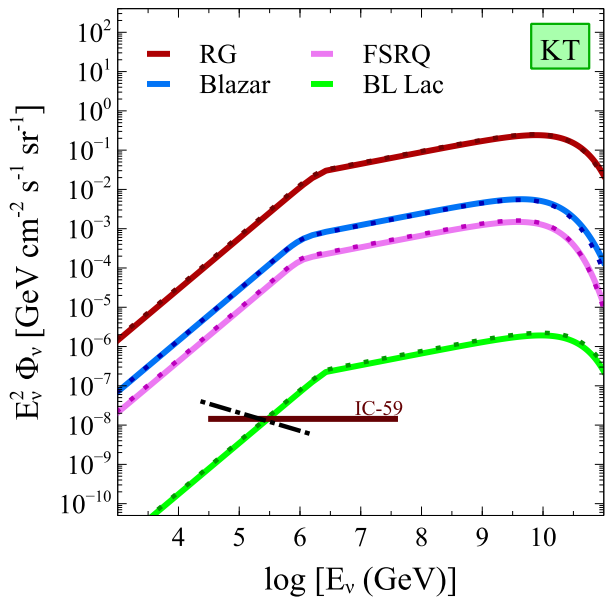}}
\caption{Predicted observed neutrino emission from various AGN source populations, total diffuse emission from contributions binned in redshift (solid line) and luminosity (dotted line). The horizontal solid line is the {\it IceCube} 1-yr (IC-59) neutrino detection limit \citep{AAA2014b}. The neutrino emission expected from the RG population (red lines) gives the highest detection, followed by emission from blazar-derived sources (green lines) and FSRQ-derived sources (blue lines). The emission from the BL Lac derived population is the only source emission we cannot definitively exclude within this model. The dash-dotted line corresponds to the {\it IceCube} best-fitting diffuse neutrino spectrum \citep{AAA2015}, where $\Phi_{\nu} = 2.06\times 10^{-18} \ [E_{\nu}/(10^{5} \ {\rm GeV})]^{-2.46} \ {\rm GeV^{-1} \ cm^{-2} \ s^{-1} \ sr^{-1}}$.}
\label{fig:kt}
\end{figure}

The resultant energy spectra expected from the KT model prescription is shown in Fig. \ref{fig:kt}, as the sum of contributions binned in source luminosity (solid line) or redshift (dashed line). 
%
%
The neutrino emission from RGs far exceeds the {\it IceCube} limit, with the dominant emission coming from local sources, $z<1$. At lower neutrino energies ($E_{\nu}\leq 10^{6}$ GeV) the greatest contribution coincides with the peak of the AGN population, i.e. $1 < z < 2$. 

The source luminosities dominating the neutrino emission are between $43.0 < \log L_{\rm X} < 45.0$. Low-luminosity sources contribute comparably, due to their abundance in the near Universe. The brightest sources are few,
   and thus only contribute significantly at lower redshifts,
   due to propagation effects.

The energy spectra of neutrinos originating in the blazar and FSRQ populations also exceed the {\it IceCube} limit. The dominant epoch is $2 < z <4$, which coincides with the peak activity of these sources. At the highest neutrino energies the local epoch dominates, and the dominant contribution at lower neutrino energies extends up to a redshift $z < 6$. The low-energy trend is stronger in the spectra with FSRQ origin than that of the full blazar population, and occurs at energies $E_{\nu} < 10^{6}$ GeV. 

The luminosities of the blazar sources responsible for the majority of the neutrino emission are $45.5 < \log L_{\rm X} < 46.5$. The lower luminosity for the FSRQ population is $\log L_{\rm X} = 46.0$, and the neutrino contribution decreases with increasing luminosity, following the same trend as the brightest blazars ($\log L_{\rm X} > 46.0$).

BL Lacs evolve negatively with redshift, and the neutrino flux from these sources follows this trend. The dominant redshift contribution is from the local epoch, with the flux decreasing with increasing redshift. The source luminosity contributions are dominated by the low-luminosity sources, and the neutrino emission similarly decreases with increasing luminosity. 

The X-ray selected BL Lacs are scarce, and the neutrino emission produced in these sources falls below the {\it IceCube} limit at lower neutrino energies. The only source population we cannot definitively exclude within the KT model prescription is therefore the BL Lacs. 

\subsubsection{BB model spectra}
\begin{figure*}
\begin{minipage}{15cm}
\centering
\resizebox{1\hsize}{!}{\includegraphics[angle=0]{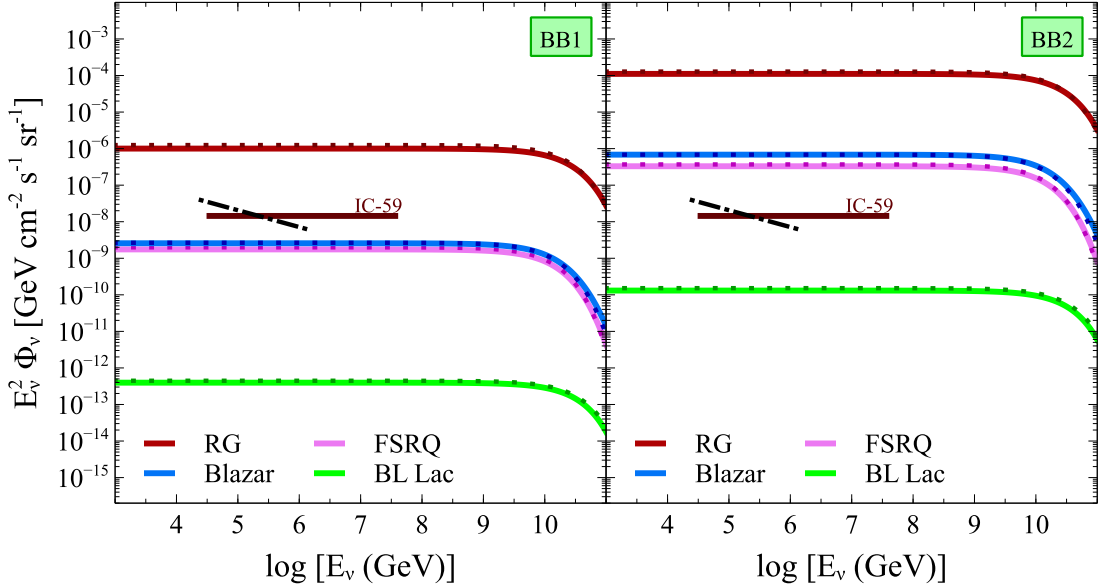}}  
\caption{Predicted observed neutrino emission from various AGN source populations, total diffuse emission from contributions binned in redshift (solid line) and luminosity (dotted line). The horizontal solid line is the {\it IceCube} 1-yr (IC-59) neutrino detection limit \citep{AAA2014b}.The neutrino emission expected from the RG population (red lines) gives the highest detection, followed by emission from blazar-derived sources (green lines) and FSRQ-derived sources (blue lines). The left panel uses the luminosity scaling BB1, which leads to the acceptance of all blazars as candidate neutrino sources. The BB2 model excludes all but the BL Lac population. The dash dotted line corresponds to the {\it IceCube} best-fit diffuse neutrino spectrum \citep{AAA2015}, where $\Phi_{\nu} = 2.06\times 10^{-18} \ [E_{\nu}/(10^{5} \ {\rm GeV})]^{-2.46} \ {\rm GeV^{-1} \ cm^{-2} \ s^{-1} \ sr^{-1}}$.}
\label{fig:bb}
\end{minipage}
\end{figure*}

The resultant energy spectra from the BB model prescription is shown in Fig. \ref{fig:bb}, for the two scaling models used. The sum of contributions binned in source luminosity (solid line) or redshift (dashed line) are in agreement, and the resultant spectra from the two scaling models emphasize the importance of the luminosity scaling. The neutrino spectra from the BB2 model are two orders of magnitude greater than those from the BB1 model. Overall, the BB model prescription produces lower expected total diffuse neutrino emission than that of the KT model. 

The RG population is seen to again greatly exceed the {\it IceCube} limit. The dominant contribution is from local sources, decreasing with increasing redshift, and the bulk of the emission originates in bright AGN, with luminosities $44.0 < \log L_{\rm X} < 46.0$. Both scaling models agree on that behaviour, however the BB1 model favours the brightest AGN over the low-luminosity sources, whereas this trend is weaker in the BB2 model. 

The effect of the luminosity scaling is significant for the neutrino emissions of blazars. The neutrino flux from blazars falls below the {\it IceCube} limit using the BB1 scaling, with the dominant contribution from AGN with $45.5 < \log L_{\rm X} < 47.0$. The BB2 scaling, however, finds comparable contributions from lower luminosity sources ($44.0 < \log L_{\rm X} < 45.5$) and brighter sources ($46.5 < \log L_{\rm X} < 47.0$), and the total diffuse emission exceeds the {\it IceCube} limit by about two orders of magnitude. 

The epoch enclosing the peak activity of blazars and FSRQs results in the majority of the diffuse neutrino flux in these sources, $2 < z < 4$. As the FSRQs are already assumed to be the upper end of the blazar luminosities, the neutrino emission decreases with increasing FSRQ luminosity. However, the flux from the epoch of $4 < z < 6$ is slightly higher relative to the blazar population, due to the lower luminosities tend towards lower redshifts, and are therefore not found in the FSRQ population.

The neutrino spectra from the BL Lac population reflect the source evolution, similarly to the KT model spectra. As the population evolves negatively with redshift, the dominant neutrino contribution originates at the lowest source luminosity, and in the most local sources. The BL Lacs are the only population to be accepted as a possible neutrino producing source within the model prescriptions considered above. 

\section{Summary and Conclusion}\label{sec:implications_discussion_conclusion} 
We have calculated the XLF, and derived the total numbers, for RL AGN and blazars in different luminosity and redshift ranges. These AGN demographics are then convolved with the neutrino production model to obtain the muon neutrino energy spectra as detected on Earth. It is apparent that the neutrino emission received from the epoch of AGN peak activity is enhanced, which is a consequence of the X-ray luminosity scaling. Additionally, the importance of the luminosity scaling model is highlighted by our results (Fig. \ref{fig:bb}).

We test a number of assumptions in these calculations. The AGN source distributions are assumed to derive from complete surveys, and the evolution is correctly represented by the XLFs \citep{SGB2008,ACS2009}. This gives us a fair estimate of the AGN populations at different redshifts, and correct proportions over the range of X-ray luminosities. 

The modest bulk Lorentz factor, $\Gamma = 10$ is assumed typical for all AGN. It provides a correction factor for blazars accounting for misaligned sources, $\Theta_{\rm CF} = 400$, which agrees with estimates in literature \citep[e.g.][]{RM1998}. It also determines the beaming to intrinsic source luminosity relation for blazars, which affects in the luminosity scaling of the resultant neutrino spectra. Extreme blazars can have $\Gamma > 20$ \citep{M2009}, but $<30$ \citep{RM1998} thus future studies would do well in exploring the effects of varying Lorentz bulk factors of AGN jets on the resultant neutrino emission.

We assume a correction for undetected sources in the RG population following \citet{ZMI2011}, assuming the detected fraction is 10\% of the total population. As we want to study RGs, we assume the RL fraction of the survey is 10\% \citep{UP1995}, thus yielding a correction factor of $\Theta_{\rm CF}$. 

Due to the scarcity of the BL Lacs population \citet{ACS2009} note that they can only fit a single power-law LF. This may reflect an unfair representation of the neutrino emission from this sub-population of blazars. Our neutrino emission calculations are only as good as the source density model, and any conclusion drawn with respect to neutrino emission from low or high luminosity AGN, may not be valid. It is therefore worth improving the survey data to lower luminosities to fill in the lower end of the LF, as well as towards higher redshifts. It would also be interesting to explore the intrinsic XLF of the blazar populations \citep[e.g.][]{ASR2012,ARG2014}.

The modelling of neutrino production in AGN jets relies on the knowledge of particle interaction channels, and the branching ratios of the interaction. The production site is assumed to be at the base of the AGN jet as the environment in the vicinity of the black hole core is sufficiently energetic for high-energy particle interactions to occur. Models of high-energy neutrino output from AGN lobes, the torus or intergalactic media, due to $pp$ or $p\gamma$ interactions are also worth studying \citep[e.g.][and references therein]{BB2009,R2011}. 

The effect of the luminosity scaling model used in each case is shown to be crucial for the estimated neutrino spectra (Figs. \ref{fig:kt} and \ref{fig:bb}). We use a model that will favour neutrino emission from bright sources, and suppress emission from the abundant lower luminosity counterparts. The KT scaling model assumes a linear relation between the source X-ray luminosity and neutrino emission, and is normalized to that of Cen A, which is assumed to be a typical neutrino producing AGN. The BB1 and BB2 scaling models rely on the relation between the jet synchrotron and total power \citep{FMB1995,BBR2005,CMN2010}, and the jet-disc symbiosis model \citep{FB1995}, which therefore enables a relation between neutrino emission in the jet to the X-ray disc luminosity. 

The resultant diffuse neutrino fluxes predicted using these models exceed the observational flux limit set by \citet{AAA2014b}, implying the following: \\ 
(i) Cen A might not be a typical neutrino source as commonly assumed. If Cen A is an exceptionally efficient neutrino machine \citep{KT2008} the scaling of the neutrino yield will lead to an overestimated neutrino flux, similar to what our calculations show (Figure \ref{fig:kt}). \\ 
(ii) The two neutrino flux models we considered have overestimated the neutrino production efficiency. The KT and BB models are motivated by an observed correlation between UHECRs and local AGN \citep{PAO2007,PAO2008}. The models assume a correlation between CR and neutrino emissions due to their common production path, e.g. through $p\gamma$ interactions. The KT model is based on the observation of UHECR emission originating in the vicinity of Cen A, however \citet{LW2009} discuss the possibility of an accidental correlation between Cen~A and the observed UHECR events.  \\ 
(iii) Neutrino luminosity does not universally scale with the accretion power for all AGN subclasses, 
 and hence not with their X-ray luminosity. 
This will require a more complex class dependence scaling prescription than the simple universal scaling that we have used here. The three scaling models we use are linear (KT) or power laws (BB1, BB2). The steepest scaling is given by the BB1 model, and is seen to suppress the contribution from the abundant lowest luminosity sources. As these sources are predominantly found in the nearby Universe, a suppression will then enhance the neutrino contribution from the epoch coinciding with the peak activity of bright AGN, at redshifts $2 < z < 4$. \\  
(iv) Some AGN are not neutrino sources. For instance, there could be a power threshold only above which charged particles could be accelerated efficiently and neutrino production could occur. The low-luminosity FR~Is may not be sufficiently powerful for the acceleration of particles to energies of $10^{20}$ eV \citep{LW2009}. If FR~Is are the parent population of BL Lac sources, then this would also apply to this blazar subclass. Energy loss calculations of the Cen A jet \citep{RMR2011} find that Cen A is unable to obtain a proton energy exceeding $E_{p,{\rm max}} \sim 10^7~{\rm GeV}$.  This is supported by indications of lower Lorentz factors in FR~Is than FR~II \citep{DMI2014}. If FR~IIs are the parent population of FSRQs, then a neutrino correlation with CRs may be weak or negligible, as FR~IIs are unfavoured as UHECR producers \citep{KO2010}. A highly efficient jet environment could lead to the UHECR population decaying before escaping the confinement, hence only neutrinos would be observable. \\ 
(v) Neutrino generation and X-ray emission have different duty cycles. Jets may have alternating duty cycle of baryonic and lepton flows, or neutrino production could occur only during some fraction of the entire X-ray lifetime of the AGN. \\ 
(vi) It is a combination of some of the above.  

\section*{Acknowledgements}
We thank Drs Kumiko Kotera and Ranieri Baldi   
for helpful discussion and comments. 
AYLO's research  is supported by the Brunei Government Scholarship. 
IBJ acknowledges the support from  L\aa nekassen, Norway.
This research has made use of NASA's Astrophysics Data Systems. 

\bibliographystyle{mn2e}
\bibliography{jour,references}

\label{lastpage}

\end{document}